# What is a "bug"?

On subjectivity, epistemic power, and implications for software research

**David Gray Widder**
Digital Life Initiative, Cornell University, New York City.

**Claire Le Goues**
School of Computer Science, Carnegie Mellon University, Pittsburgh.

> *"Perception always intercedes between reality and ourselves"* [1]

Considerable effort in software research and practice is spent on bugs. Finding bugs, reporting and tracking them, triaging them, attempting to fix them automatically, detecting "bug smells" — these comprise a substantial portion of large projects' time and development cost, and are a significant subject of study for researchers in Software Engineering, Programming Languages, Security, and beyond.

But, what *is* a bug, exactly? While segmentation faults rarely spark joy, most bugs are not so clear cut. Per the Oxford English Dictionary, the word "bug" has been a colloquialism for an engineering "defect" at least since the 1870s. Most modern software-oriented definitions – formal or informal – speak to a disconnect between what a developer intended and what a program actually does. Formal verification, from its inception, has developed means to identify deviations from a formal specification, expected to more or less fully encode desired behavior. However, software is rarely accompanied by full and formal specifications, and this intention is instead treated as implicit or partially-documented at best. For example, the International Software Testing Qualifications board writes: "A human being can make an error (mistake), which produces a defect (fault, bug) in the program code, or in a document. If a defect in code is executed, the system may fail to do what it should do (or do something it shouldn't), causing a failure. Defects … may result in failures, but not all [do]" [2]. Most sources forsake even that much precision. Indeed, The influential paper "Finding bugs is easy" begins by saying "bug patterns are code idioms that are often errors" — with no particular elaboration [3]. Other work relies on imperfect practical proxies for specifications. For example, in automatic program repair research, a bug corresponds to a failing test case: when the test passes, the bug is considered fixed.

However, when we interrogate fairly straightforward definitions, they start to break down. Simple corner cases abound: developers often encode non-functional checks using style checkers, but we'd be hard-pressed to declare an incorrect comment indentation a "bug." Meanwhile, even tests that unambiguously check desired functionality can fail for reasons wholly unrelated to correctness. These "flaky tests" sometimes pass and sometimes fail, due to problems like race conditions or resource contention – but are they bugs?

Looking beyond specifications or corner cases, subjectivity and judgment surrounds what a bug is, and how and whether it should be fixed. There are almost always more bugs reported than can be reasonably handled given available resources, a fact baked into modern continuous-deployment pipelines and bug triage processes. Or, consider *error budgets*, which help balance business risk against development velocity, premised on an acceptance that software systems *inevitably* contain errors, because it is not cost effective to fix them all.

The fact that software ships with bugs bakes into modern engineering a continuous set of judgements about which bugs must be fixed. This regular use of *judgment* concerning is based on the *subjectivity* of the person exercising this judgment. Researchers have studied these judgments, looking for example at the factors that appear to influence which bugs get fixed, how quickly, and by whom. Many of these factors are *not* purely technical, but reflect subjective interpretation, relative to the software creator's goals. User-submitted bug reports are categorized as INVALID or WONTFIX fairly often, to reflect behavior that either cannot be changed or that corresponds to developer-intended behavior [4]. Even while requirements engineering has long recognized that building software entails possibly contradicting viewpoints [5], a review of requirements engineering in Agile software development—which emphasizes frequent interaction with one's intended users—found that "building a shared understanding of the user perspective is not very well established" [6]. Being experts on their software, project maintainers may be in a position to make this judgment call! Importantly, however, this doesn't make that judgment call *objective*. This influences how software engineering research proceeds.

*Epistemology* is the branch of philosophy that deals with how we make and justify claims of truth. All research and scholarship adopts an epistemological stance, though often tacitly. When we conduct Science, we often adopt some form of *positivism*. Roughly, this view asserts that there is one objective reality that we all can observe or measure, and that results we find are obvious such that we can easily agree on what they are. However, sociologists examining the process of creating new science show how science is also a process of persuading reviewers, peers, politicians, and the public [7]. Indeed, the way we scientists interpret our own and others' results is informed by our subjectivity: beliefs, values and other aspects of our lived experiences. This is closer to an epistemological stance called *constructivism*: there might not be one "reality" that everyone experiences and makes sense of the same way, but that instead, the way we make sense of our world is an inherently social process, constructing meaning in dialogue with others. Even Scientific "Truth" is thus socially situated. In this view and ours, what a "bug" is, is contextual and dependent on who is empowered to decide.

This highlights the role of *Epistemic Power:* the power to produce knowledge, and thereby influence what others think, believe, and accept as true [8]. Researchers can attempt to use their epistemic power to center other subjectivities besides what might appear to be the default. We illustrate with two bug-based examples:

- GenderMag [9] is a tool to fix "gender inclusiveness bugs", by helping find nuances in software design that make it harder to use by, for example, women, based on research showing how problem solving strategies tend to cluster by sex[1]. GenderMag attempts to allow software designers–typically men–to adopt the subjectivity of someone who uses software differently than they do, to help them design more widely usable–and thus better– software. Notably, these design nuances are positioned squarely as "bugs" to acknowledge the range of subjective experiences that might impede someone using otherwise "functional" software. This uses the moral weight of "bug" to motivate software teams to fix issues they might not otherwise consider in-scope.

- A recent survey[2] of 115 software engineers' ethical concerns they had encountered in their work revealed an extraordinary range [10]: from a numerical *bug* that could potentially kill crane operators, to concerns about the ethics of a business altogether (when working for a military contractor, for example). Many engineers found it difficult to secure support to fix even straightforward bugs, because doing so would be "too much work", or that "the client didn't ask for it". Here, the manager or client gets to determine what ethics issues count as "bugs" worth fixing.

GenderMag makes software more inclusive by expanding the definition of bug, and demonstrates that hunting for new kinds of bugs can enable a broader view of design; The ethics survey shows how the power to define bug has material ethical effects. Together, they show that when we critically examine what a bug is, interesting research opportunities arise.

---

[1] The GenderMag paper may in places conflate gender, after which the tool is named, with sex, the construct used in the underlying research it is based on. See discussion of the two constructs in the paper.
[2] David Gray Widder is an author of the present piece, and of the referenced study.

The "what is a bug?" question reveals the limits of a positivist approach to research. Bugs become meaningful only when people agree, and this agreement isn't certain. Often, those with various kinds of power (clients with money, people with software engineering knowledge) scaffold this agreement. At a higher level, software research and practice are conducted in the context of a society where the effects of power, and of divergent interests, lead some to benefit from software and others to be harmed by it, disproportionately. Leveling the effects of epistemic power broadly requires wider systemic and societal changes, which we must all play a role in. But researchers and engineers, even those of us focused purely on conventional software quality and development concerns must also be cognizant of how this affects our research and practice.

First, new software development techniques involve design choices that are intrinsically grounded in subjective experiences. Technique designers (like SE researchers) should include and justify these judgment calls in scoping scientific claims, including how generalizability may be affected. Mulligan and Bamberger advocate for "contestable design" exposing value-laden choices to enable "iterative human involvement in system evolution and deployment" [11]. By making it more transparent where judgment calls were made in the science we do, scholars (especially new ones) will find it easier to read papers as a series of judgment calls and established norms, rather than as the natural or obvious ways to do things. This will also make it easier for scholars with different subjectivities to better understand (and then explore) how things could proceed differently, leading to a richer and more robust scientific process.

This type of scoping will benefit from an attitudinal change in peer review, since it motivates methods sections that explain why something was done in the face of other reasonable alternatives (not just bracketing off alternatives to a bulletproof "threats to validity" section). Instead of seeking cracks in a paper's armor, reviewers ought to demand these "cracks" be exposed plainly (and indeed not see them as cracks), to enable continued contestation and experimentation. This lies in contrast with common advice given to junior scholars to reduce the "attack surface" of a paper, the exact opposite of contestability.

Second, our science should ground its aims and results in the needs and concerns of humans. In research that aims to find and fix bugs, it remains fair to check performance on historical bug datasets, so long as we don't lose sight of the people that improved software quality is intended to help. This likely means more, and more carefully considered human-in-the-loop studies, and more reviewer acceptance of alternative modalities for the use and evaluation of proposed techniques beyond their performance on historical bug datasets. That said, this should be done

carefully — some work relies heavily on the fact that open source project maintainers have accepted bug reports or patches produced by some new technique, without interrogating or considering the social factors that influence that acceptance, which we discuss above.

That said, advocating for "human in the loop" research begs the question: which humans are we centering? Research in bug finding and fixing often uses databases of bugs drawn from open source repositories or similar databases, which correspond to issues that project maintainers understood well and declared sufficiently important to fix. Assumptions are therefore baked into those historical datasets. What if a bug was deprioritized for triage and repair because it affects a smaller user group, but that group is disproportionately disabled, or female? This might motivate an effort to collect bugs from more diverse user groups, and opportunities to discuss differences found through this process in research papers.

The GenderMag and "Ethics Bugs" examples illustrate the value of thinking more widely about what is considered a bug. Some work already does this, looking at quality attributes like performance or security. Ethical or human-centered concerns can rise to the same level of consideration. For example, in Pittsburgh, sidewalk food delivery robots waited on sidewalk curb ramps for crosswalk signals to change, stranding wheelchair users in traffic lanes [12]. This is surely a bug, as worthy of inclusion in historical bug datasets as a more traditional segmentation fault.

**Acknowledgements**